\documentclass[12pt,thmsa,arabtex]{article}
\usepackage{amsmath}
\usepackage{amssymb}
\usepackage{graphicx}

\input{tcilatex}

\begin{document}

\begin{center}
\textbf{\large Qualitative aspects of the entanglement in the three-level
model with photonic crystals}

~

Mahmoud Abdel-Aty\footnote{%
E-mail: abdelatyquantum@yahoo.co.uk, Fax. No. 00-20-93-4601950}

{\small Mathematics Department, Faculty of Science, South Valley University,
82524 Sohag, Egypt. }
\end{center}

\textbf{Abstract:}

{\ This communication is an enquiry into the circumstances under which
concurrence and phase entropy methods can give an answer to the question of
quantum entanglement in the composite state when the photonic band gap is
exhibited by the presence of photonic crystals in a three-level system. An
analytic approach is proposed for any three-level system in the presence of
photonic band gap. Using this analytic solution, we conclusively calculate
the concurrence and phase entropy, focusing particularly on the entanglement
phenomena. Specifically, we use concurrence as a measure of entanglement for
dipole emitters situated in the thin slab region between two semi-infinite
one-dimensionally periodic photonic crystals, a situation reminiscent of
planar cavity laser structures. One feature of the regime considered here is
that closed-form evaluation of the time evolution may be carried out in the
presence of the detuning and the photonic band gap, which provides insight
into the difference in the nature of the concurrence function for atom-field
coupling, mode frequency and different cavity parameters. We demonstrate how
fluctuations in the phase and number entropies effected by the presence of
the photonic-band-gap. The outcomes are illustrated with numerical
simulations applied to GaAs. Finally, we relate the obtained results to
instances of any three-level system for which the entanglement cost can be
calculated. Potential experimental observations in solid-state systems are
discussed and found to be promising. }

\bigskip


\bigskip

\section{Introduction}

A structure in which the dielectric constant varies periodically is called a
photonic crystal. One of the most interesting properties of a photonic
crystal is the existence of a photonic band gap \cite{lee04,yam03}.
Radiation with a frequency that lies within the band gap cannot propagate in
the photonic crystal structure. Photonic crystals are usually viewed as an
optical analog of semiconductors that modify the properties of light similar
to a microscopic atomic lattice that creates a semiconductor band-gap for
electrons \cite{lee04,kam00,joa95}. Photonic band gap crystals offer unique
ways to tailor light and the propagation of electromagnetic waves and have
caused growing interest in recent years because it offers the possibility of
controlling and manipulating light within a given frequency range through
photonic band gap \cite{lee04,kam00}. Photonic band-gap materials have
attracted much attention in recent years for theoretical and practical
importance in fundamental science and application \cite{lee04,joa95}. The
atom-photon interaction in photonic band gap materials \cite{joa95} has been
found to exhibit many interesting new phenomena such as photon-atom bound
states \cite{joh90}, spectral splitting \cite{joh94}, quantum interference
dark line effect \cite{zhu97}, phase control of spontaneous emission \cite%
{pas98}, transparency near band edge \cite{pas99}, and single-atom switching 
\cite{flo01}.

In a parallel development, considerable work was done recently on
entanglement properties \cite{vid03}. The detection of entanglement is one
of the fundamental problems in quantum information theory. From a
theoretical point of view one can try to answer the question whether a given
entirely known state is entangled or not, but despite a lot of progress in
the last years [11,12], no general solution of this problem is known. In
experiments, one aims at detecting entanglement without knowing the state
completely. Bell inequalities \cite{per99} and entanglement witnesses \cite%
{hor96} are the main tools to tackle this task. Interestingly, the
concurrence of the ground state which is related to the entanglement of
formation, has been shown to be strongly affected at the critical point \cite%
{woo98}. More precisely, in the one-dimension, it has been shown that the
derivative of the concurrence with respect to the coupling constant diverges
at the transition point, although the concurrence itself is not maximum.
These pioneering results raise the question of the universality of these
behaviors. Actually, the lack of exact solutions especially in higher
dimensions implies a numerical treatment which often restrict the study to a
small number of degrees of freedom.

Heisenberg's uncertainty relations had tremendous impact in the field of
quantum optics particularly in the context of the construction of coherent
states and also for different physical systems as well as the reconstruction
of quantum states. The minimization problem of finding the number-phase
uncertainty state has been considered and minimum uncertainty state
relations between number and phase uncertainty are presented \cite{bia75}.
Many authors argued that \cite{buz98}, the Heisenberg inequality is too weak
for practical purposes, which led them to the establishment of information
theoretic uncertainty relations.

Our aim of the present paper is to consider the dynamics of a system of
three-level atoms with dipole interaction in presence of the photonic band
gap and study the concurrence and the entropic uncertainty relation for
number and phase. With applying some approximations, one can deal with the
quantization of the electromagnetic field modes of a homogeneous, but
anisotropic medium which can then be made to form one of the sandwich layers
in the slab structure under consideration involving two semi-infinite
periodic photonic crystals. With the electromagnetic modes quantized, one
can evaluate the entanglement degree, and explore its variations with the
controllable parameters of the system. To reach our goal we have to find an
exact analytic solution of the time dependent Schr\"{o}dinger equation of
the system. We show that a reasonable amount of entanglement can be achieved
in a system of three-level atoms with dipole interaction in presence of the
photonic band gap and essentially we establish deeper connections between
entropic uncertainty relations and entanglement.

The organization of this paper is as follows: in section 2, we give an
overview of effective medium approach and dispersions, followed by
subsection 2.1 where we introduce our Hamiltonian model and give exact
analytic solution for the Schr\"{o}dinger equation in the frame of the
dressed state formalism. In section 3, we employ the analytical results
obtained in section 2 to investigate the properties of the entanglement
degree due to the concurrence, and classify the behavior in several
parameter regimes assuming that the electromagnetic field is in a coherent
state in subsection 3.1. In section 4, we essentially establish deeper
connections between entropic uncertainty relations and entanglement.
Numerical results for the phase entropy are discussed in the subsection 4.1
for two different cases; one is the resonant and the other is the
off-resonant case. The prospects for experimental observation of our
predictions are analyzed in section 5. Finally, a summary of the main points
of this work ends the paper and a few avenues for further investigations are
indicated in section 6.

\section{Effective medium approach}

The effective-medium approach can be applied to situations in which all
three regions of the structure possess frequency-dependent dielectric
functions. In fact, the rapid pace of the technological progress in
solid-state quantum computing gives one a hope that the specific
prescriptions towards building robust qubits and their assemblies discussed
in this work can be implemented in future devices. In this regard, very
promising fields where the concept of nonlinear localized modes may find
practical applications is the quantum computation of photonic band gap
materials, periodic dielectric structures that produce many of the same
phenomena for photons as the crystalline atomic potential does for electrons
[4,18]. Nonlinear photonic crystals (or photonic crystals with embedded
nonlinear impurities) create an ideal environment for the generation and
observation of nonlinear localized photonic modes. Much theoretical work has
been done on the properties of finite one dimensional photonic band gap
(PBG) crystals \cite{joa95}, including recent calculations of the thermal
emissivity of such one-dimensional structures \cite{cor99}. The strong
angular dependence of the gap effect with a one-dimensional structure has
motivated successful experimental work with three-dimensional structures 
\cite{lin00}. In particular, the existence of such modes for the frequencies
in the photonic band gaps has been predicted \cite{joh93} for $2D$ and $3D$
photonic crystals with Kerr nonlinearity. Nonlinear localized modes can also
be excited at nonlinear interfaces with quadratic nonlinearity \cite{suk99},
or along dielectric waveguide structures possessing a nonlinear Kerr-type
response \cite{mcg99}.

\begin{figure}[tbph]
\begin{center}
\includegraphics[width=9cm]{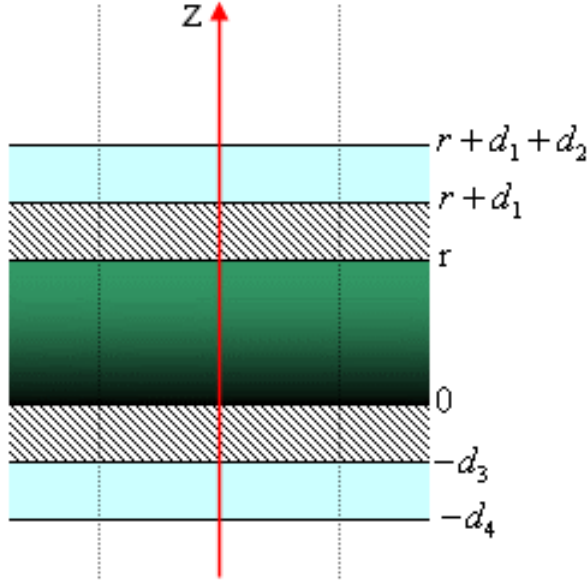}
\end{center}
\caption{A schematic representation of the dielectric slab between two
photonic crystals occupying the regions $z<0$ and $z>r$.}
\end{figure}
The system that we consider here consists of a dielectric cavity occupying
the region $0<z<r$ and the photonic crystals occupy the regions $z>r$ and $%
z<0$. For long wavelength fields and in the effective medium approach, the
photonic crystal has the optical characteristic of a uniaxial medium \cite%
{kam00}. Moreover we shall specialize to uniaxial media, so that our system
has only two principal axes with the $z-$axis as the optical axis. In this
case, the components of the dielectric tensor appropriate to the photonic
crystals can be written as

\begin{equation}
\epsilon =\left( 
\begin{array}{c}
\epsilon \\ 
0 \\ 
0%
\end{array}
\begin{array}{c}
0 \\ 
\epsilon \\ 
0%
\end{array}
\begin{array}{c}
0 \\ 
0 \\ 
\epsilon _{z}%
\end{array}
\right) ,  \label{co}
\end{equation}
where, $\epsilon =\epsilon _{0}\epsilon ^{||},$ $\epsilon _{z}=\epsilon
_{0}\epsilon _{z}.$ The dielectric tensor components for the two
semi-infinite crystals can be written in the following forms $\epsilon
_{1}^{||}=(\eta _{1}d_{1}+\eta _{2}d_{2})/d_{12},\qquad \epsilon _{z1}=\eta
_{1}\eta _{2}d_{12}/(\eta _{1}d_{2}+\eta _{2}d_{1}),$ and $\epsilon
_{2}^{||}=(\eta _{3}d_{3}+\eta _{4}d_{4})/d_{34},\qquad \epsilon _{z2}=\eta
_{3}\eta _{4}d_{34}/(\eta _{3}d_{4}+\eta _{4}d_{3}),$ where the $
d_{ij}=d_{i}+d_{j},$ the subscripts 1 and 2 on $\epsilon _{i}^{||}$ and $
\epsilon _{zi}$ refer to the first and second photonic crystal. The
dielectric functions $\eta _{1}$ and $\eta _{2},$ one or both of which may
be frequency dependent. The photonic crystals are treated using the
effective medium approach, which pertains to any layer structure formed by
alternate periodic stacking of two types of layers of locally isotropic
materials of thicknesses d$_{1}$ and $d_{2}$ (see figure 1).

In this paper we are concerned with the interface polaritons which are
characterized by imaginary wave vectors normal to the interfaces such that
the waves are decaying with distance from the interfaces at $z=0$ and $z=r$
into the outer regions and are hyperbolic in the slab \cite%
{kam00,mcg99,cot89}. To see the salient features of the effective medium
description we shall ignore retardation effects, which amounts to ignoring
throughout ($\omega /c$) terms. In this case dispersion relation for the
surface polaritons takes the form

\begin{equation}
k_{s}r=\arctan h\left( -\frac{k_{s}}{\epsilon _{s}}\times \frac{
(k_{1}/\epsilon _{1}^{||})+(k_{2}/\epsilon _{2}^{||})}{(k_{s}/\epsilon
_{s})+(k_{1}k_{2}/\epsilon _{1}^{||}\epsilon _{2}^{||})}\right) .
\end{equation}
The dispersion relations obtained from the Maxwell wave equation of this
system lead to two distinct equations \cite{kon75} $k_{s}^{2}=k_{||}^{2}-
\omega ^{2}\epsilon _{s}/c^{2},\quad k_{i}^{2}=\epsilon
_{i}^{||}k_{||}^{2}/\epsilon _{zi}-\omega ^{2}\epsilon _{i}^{||}/c^{2},$
where s refers to the slab cavity.

It is important to note that infinite and semi-infinite photonic crystals
have the same band structure \cite{zol98}. The only difference is the
existence of surface modes in the case of semi-infinite structure. The main
feature of all 1D photonic crystals is that although forbidden gaps exist
for most given values of the tangential component of the wave vector ($k$),
there is not an absolute nor complete photonic band gap if all possible
values of the tangential component of the wave vector are considered \cite%
{joa95}. Having determined the modes we can now quantize the fields
associated with these modes using the usual quantization procedure \cite%
{kam00} the single-mode quantized field takes the form 
\begin{equation}
E(\hat{x},t)=E_{0}\hat{a}(\hat{k}_{||})\exp [i(\hat{k}_{||}.\hat{x}-\omega
t)]+H.C.,
\end{equation}
where $E_{0}$ is the strength of the electric field, $\hat{k}_{||}$ is the
wave vector, $\hat{x}$ is the position operator and $\hat{a}$ the
annihilation operator.

\subsection{The model and methods of solution}

Accurate potentials are of course required for a quantitatively correct
prediction of the behavior and properties of real quantum systems. However,
even qualitative conclusions drawn from simulations employing inaccurate or
invalidated potentials can be problematic. The most appropriate form of the
potential depends largely upon the properties of interest to the simulators.
Now we consider the interaction of the abovementioned modes with a
three-level atom in three different configurations, namely, $V-,$ Lambda-
and cascade-type. The transition in the 3-level atom is characterized by the
dipole matrix element $\lambda _{ij}.$ The operator $\hat{S}_{ii}$ describes
the atomic population of level $|i\rangle _{A}$ with energy $\omega
_{j},(j=a,b,c)$ and the operator $\hat{S}_{ij},(i\neq j)$ describes the
transition from level $|i\rangle _{A}$ to level $|j\rangle _{A}$. 
\begin{figure}[tbph]
\begin{center}
\includegraphics[width=15cm]{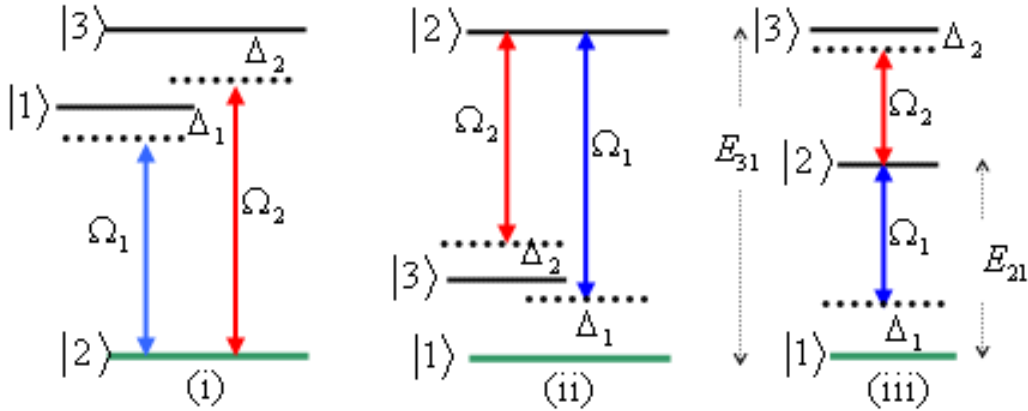}
\end{center}
\caption{The V-type, $\Lambda $-type and $\Xi $-type three-level atom
interacting with a bimodal field. The levels $|1\rangle _{A}$, $|2\rangle
_{A}$, and $|3\rangle _{A}$, have the energy values $\hbar \protect\omega %
_{1},\hbar \protect\omega _{2}$ and $\hbar \protect\omega _{3}$,
respectively. The transitions $|1\rangle _{A}\longrightarrow |2\rangle _{A}$
, and $|2\rangle _{A}\longrightarrow |3\rangle _{A}$, are coupled to two
intra-cavity different modes $\hat{a}_{1}$ and $\hat{a}_{2}$ with
eigenfrequencies $\Omega _{1}$ and $\Omega _{2}$. The detunings of the
levels $|1\rangle _{A},|2\rangle _{A}$, and $|2\rangle _{A},|3\rangle _{A}$,
are $\Delta _{1}=\protect\omega _{1}-\protect\omega _{2}-\Omega _{1}$ and $%
\Delta _{2}=\protect\omega _{3}-\protect\omega _{2}-\Omega _{2}$, for
V-type, and $\Xi $-type, while $\Delta _{1}=\protect\omega _{2}-\protect%
\omega _{1}-\Omega _{1}$ and $\Delta _{2}=\protect\omega _{2}-\protect\omega %
_{3}-\Omega _{2}$ for $\Lambda $-type. }
\end{figure}
The total Hamiltonian of this system is $\hat{H}=\hat{H}_{0}+\hat{H}_{int}$.
The $3$ eigenstates, $\left\vert \xi _{i}\right\rangle $ and corresponding
eigenenergies, $\alpha _{i}$ are assumed to be known. The total
wave-function may be expanded in terms of the known eigenstates, namely 
\begin{equation}
\left\vert \Psi (t)\right\rangle =A_{1}(t)\left\vert \xi _{1}\right\rangle
+A_{2}(t)\left\vert \xi _{2}\right\rangle +A_{3}(t)\left\vert \xi
_{3}\right\rangle .  \label{wEq}
\end{equation}%
With atomic units, using Schr\"{o}dinger equation, we obtain the coupled
equations for our three-level system, namely 
\begin{equation}
i\frac{\partial A_{j}(t)}{\partial t}=r_{j}A_{j}(t)+\sum%
\limits_{k=1}^{3}H_{jk}A_{k}(t),  \label{Pro}
\end{equation}%
where $\hat{H}_{0}\left\vert \xi _{i}\right\rangle =r_{i}\left\vert \xi
_{i}\right\rangle $ and $H_{jk}=\left\langle \xi _{j}\right\vert \hat{H}%
_{int}\left\vert \xi _{k}\right\rangle .$ These equations are exact for any
three-level atom. In the interaction picture, let us consider a three-level
system described, in an appropriate rotating frame, by the Hamiltonian 
\begin{equation}
\hat{H}_{int}=\Delta _{1}\hat{S}_{11}+\Delta _{2}\hat{S}_{33}+\lambda _{21}%
\hat{R}_{1}\hat{S}_{21}+\lambda _{32}\hat{R}_{2}\hat{S}_{32}+\lambda
_{21}^{\ast }\hat{R}_{1}^{\dagger }\hat{S}_{12}+\lambda _{32}^{\ast }\hat{R}%
_{2}^{\dagger }\hat{S}_{23}.
\end{equation}%
The atom-field couplings $\lambda _{ij}$ are given by $\lambda _{ij}=Y\mu
_{ij}.E,$where $E$ $\ \ $is the quantized electric field given by equation
(3) and $\mu _{ij}$ is the matrix dipole moment coupling between the state $i
$ and $j$. The $Y$ factor accounts for local field effects and is given by $%
Y=3\epsilon _{s}(\omega )/(2\epsilon _{s}(\omega )+1),$ where $\epsilon
_{s}(\omega )$ is given in equation (1). It is easy to write $\lambda _{ij}$
in the following form 
\begin{equation}
\lambda _{ij}=\frac{3\epsilon _{s}(\omega )}{2\epsilon _{s}(\omega )+1}.%
\frac{\left( \omega /\omega _{T}\right) ^{2}-\left( \omega _{L}/\omega
_{T}\right) ^{2}}{\left( \omega /\omega _{T}\right) ^{2}-\eta ^{2}},
\end{equation}%
where $\eta ^{2}=[$ $2\epsilon _{s}(\omega )\left( \omega _{L}/\omega
_{T}\right) ^{2}+1]/[2\epsilon _{s}(\omega )+1]$. The transitions between
the three levels may occur in three different configurations depending upon
the relationship between the energies $E_{1},E_{2}$ and $E_{3}$ of levels $%
1,2$ and $3$. The possible configurations are \cite{abd87} (i) the $V$-type
corresponding to $E_{2}<E_{1}<E_{3}$, (ii) the $\Lambda -$type or Raman
configuration corresponding to $E_{1}<E_{3}<E_{2}$ and (iii) the $\Xi -$type
or ladder-type corresponding $E_{1}<E_{2}<E_{3}$. Each of the two pairs of
levels can be coupled by only one-mode or two-mode. The field operators in
the abovementioned three types are (i) $F_{1}=\hat{a}^{\dagger },F_{2}=\hat{b%
}$ for $V$-type, (ii) $F_{1}=\hat{a},F_{2}=\hat{b}^{\dagger }$ for $\Lambda $
-type and (iii) $F_{1}=\hat{a},F_{2}=\hat{b}$ for $\Xi $-type with $\hat{a}=%
\hat{b}$ if both pairs of levels are coupled by the same mode.

In order to solve equations (5), we assume that \cite{abd87} 
\begin{equation}
G(t)=A(t)+xB(t)+yC(t),
\end{equation}%
which means that 
\begin{equation}
i\frac{dG(t)}{dt}=\left( r_{1}+v_{1}^{\ast }y\right) \left\{ A(t)+\frac{%
r_{2}x+v_{2}^{\ast }y}{r_{1}+v_{1}^{\ast }y}B(t)+\frac{v_{2}x+r_{3}y}{%
r_{1}+v_{1}^{\ast }y}C(t)\right\} ,
\end{equation}%
where $v_{1}$ and $v_{2}$ are given using equations (5) and (6). We seek $%
G(t)$ such that $i\overset{.}{G}(t)=zG(t)$. This hold if 
\begin{equation*}
y=\frac{v_{2}x+r_{3}y}{r_{1}+v_{1}^{\ast }y},\quad x=\frac{%
r_{2}x+v_{2}^{\ast }y}{r_{1}+v_{1}^{\ast }y},\quad z=r_{1}+v_{1}^{\ast }y.
\end{equation*}%
After some algebra this leads to a cubic equation which has three
eigenvalues $x_{i}(y_{i})$ which determine the $z_{i}$. There are also three
corresponding eigenfunctions $G_{j}(t)=G_{j}(0)\exp (-iz_{j}t)$, where 
\begin{equation}
G_{j}(t)=M_{j1}A(t)+M_{j2}B(t)+M_{j3}C(t),
\end{equation}%
where 
\begin{equation}
M_{ji}=\left( 
\begin{array}{c}
1 \\ 
1 \\ 
1%
\end{array}%
\begin{array}{c}
x_{1} \\ 
x_{2} \\ 
x_{3}%
\end{array}%
\begin{array}{c}
y_{1} \\ 
y_{2} \\ 
y_{3}%
\end{array}%
\right) .
\end{equation}%
Now, we express the unperturbed state amplitude $A(t),B(t)$ and $C(t)$ in
terms of the dressed state amplitude $R_{j}$ 
\begin{equation}
F_{i}(t)=\sum\limits_{j=1}^{3}M_{ij}^{-1}G_{j}(t)=\sum%
\limits_{j=1}^{3}M_{ij}^{-1}G_{j}(0)\exp (-iz_{j}t),
\end{equation}%
$F_{1,2,3}(t)=A,B,C.$ Using the above equations, we can write 
\begin{eqnarray}
A(t) &=&\frac{1}{D}\left[
(x_{2}y_{3}-y_{2}x_{3})e^{-iz_{1}t}+(x_{3}y_{1}-y_{3}x_{1})e^{-iz_{2}t}+(x_{1}y_{2}-y_{1}x_{2})e^{-iz_{3}t}%
\right] ,  \notag \\
B(t) &=&\frac{1}{D}\left[
(y_{2}-y_{3})e^{-iz_{1}t}+(y_{3}-y_{1})e^{-iz_{2}t}+(y_{1}-y_{2})e^{-iz_{3}t}%
\right] , \\
C(t) &=&\frac{1}{D}\left[
(x_{2}-x_{3})e^{-iz_{1}t}-(x_{3}-x_{1})e^{-iz_{2}t}-(x_{1}-x_{2})e^{-iz_{3}t}%
\right] ,  \notag
\end{eqnarray}%
where $D=\det
(M)=x_{1}y_{2}+x_{2}y_{3}+x_{3}y_{1}-x_{1}y_{3}-x_{2}y_{1}-x_{3}y_{2}.$ We
have thus completely determined the dynamics of a three-level system in the
presence of photonic crystal.

The picture in this case is of the three-level system in the presence of
photonic band gap and the detuning, rather than the usual picture of the
three-level Jaynes-Cummings model (JCM) system. The important point to note
here is that, using the above analytic approach, any three-level Hamiltonian
is likewise exactly solvable, with precisely similar eigenvectors and
eigenvalues that are obtained directly using equations (4) and (6). In Ref. 
\cite{bou04} an analytic approach is proposed for three-level systems, based
on the Riccati nonlinear differential equation. However, the solution
obtained is valid only in certain situations. On the other hand, our
analytic approach removed the restriction that considered in the previous
work and this solution is valid for any three-level system.

Next, we discuss a frequently encountered phenomena of particular interest
in which we define the entanglement measure of the present system.

\section{Concurrence}

Quantum entanglement has recently been attracted much attention as a
potential resource for communication and information processing \cite{ben96}
. Entanglement is usually arise from quantum correlations between separated
subsystems which can not be created by local actions on each subsystem. The
concept of concurrence originates from the seminal work of Hill and Wootters 
\cite{woo98} where the exact expression of the entanglement of formation of
a system of two qubits was derived. They showed that the entanglement of
formation, an entropic entanglement monotone, is a convex monotonic
increasing function of the concurrence.

It has been shown that the concurrence of a mixed two-qubit state, $C(\rho
_{AB})$, can be expressed in terms of the minimum average pure-state
concurrence, $C\left( \left| \psi _{AB}\right\rangle \right) $, where the
minimum is taken over all possible ensemble decompositions of $\rho _{AB}.$
So that, the concurrence is defined of a mixed state $\rho $ for $2\times 2$
quantum systems, in the following form \cite{woo98} 
\begin{equation}
C(\rho )=\max \left( \sigma _{1}-\sigma _{2}-\sigma _{3}-\sigma _{4}\right) ,
\end{equation}
where the $\sigma _{i}$ are the square roots of the eigenvalues of the
product matrix $Q$, the singular values (by convention sorted in descending
fashion), all of which are non-negative real quantities 
\begin{equation}
Q=\sqrt{\rho }^{T}\sigma _{y}\otimes \sigma _{y}\sqrt{\rho },
\end{equation}
$\sigma _{y}$ is the well-known Pauli matrix, and $\sqrt{\rho }$ is any
matrix satisfying $\sqrt{\rho }=\sqrt{\rho }^{\dagger }.$ The importance of
this measure follows from the direct connection between concurrence and
entanglement of formation $E_{f}$ 
\begin{equation}
E_{f}\left( \rho \right) =-\mu _{+}\ln \mu _{+}-\mu _{-}\ln \mu _{-},
\end{equation}
where 
\begin{equation}
\mu _{\pm }=\frac{1}{2}\left( 1\pm \sqrt{1-C(\rho )^{2}}\right) .
\end{equation}
One can prove that $\rho $ is separable if and only if the concurrence is
zero.

Let us now turn our attention to the definition of the concurrence of a pure
state \cite{run01} on a $(N\times K)-$dimensional Hilbert space $\Re =\Re
_{N}\otimes \Re _{K}.$ The \textit{flip operator }$\mathit{F}$\textit{\ }
acting on an arbitrary Hermitian operator $A$ on $\Re $ can be written as 
\begin{equation}
F(A):=A+(trA)\Pi -(tr_{N}A)\otimes \Pi _{K}-\Pi _{N}\otimes (tr_{K}A),
\end{equation}
where $tr_{N}$ and $tr_{k}$ the partail traces over $\Re _{N}$ and $\Re
_{K}, $ respectively. We denote by $\Pi _{N}$ and $\Pi _{K}$ the identity on 
$\Re _{N}$ and $\Re _{K},$ respectively. The expectation value $\left\langle
\psi \right| F(\rho _{\psi })\left| \psi \right\rangle ,$ where $\rho _{\psi
}=\left| \psi \right\rangle \left\langle \psi \right| $, is non-negative for
all pure states and equals zero if and only if $\left| \psi \right\rangle $
is a product state. This allows to define the concurrence of any arbitrary
bipartite pure state as \cite{run01} 
\begin{eqnarray}
C\left( \left| \psi \right\rangle \right) &=&\sqrt{\left\langle \psi \right|
F(\rho _{\psi })\left| \psi \right\rangle }  \notag \\
&=&\sqrt{2\left( \left\langle \psi |\psi \right\rangle ^{2}-tr(\rho
_{N}^{2})\right) },
\end{eqnarray}
where $\rho _{N}=tr_{K}\left( \rho _{\psi }\right) $ is the reduced density
operator of dimension N. For a normalized state, $\left\langle \psi |\psi
\right\rangle =1,$ it interpolates monotonously between zero for product
states and $\sqrt{\frac{2(N-1)}{N}}$ for maximally entangled states.

To investigate the concurrence for the system under consideration, we have
to evaluate the reduce atomic density matrix $\rho _{_{A}}=tr_{_{F}}\rho (t),
$ which can be written as 
\begin{equation}
\rho _{_{A}}=\sum_{i=1,2,3}\rho _{_{ii}}\left| i\right\rangle \left\langle
i\right| +\sum_{i,j=1,2,3,i\neq j}^{{}}\rho _{_{ij}}\left| i\right\rangle
\left\langle j\right| ,
\end{equation}
where $\rho _{ij}(t)=\langle i|\rho _{_{A}}(t)|j\rangle ,\quad i,j=1,2$ and $
3.$ Using equations (19) and (20), we can write the concurrence in the
following form 
\begin{equation}
C\left( \left| \psi \right\rangle \right) =\sqrt{2\sum_{i,j=1,2,3,i\neq
j}^{{}}\left( \rho _{_{ii}}\rho _{_{jj}}-\rho _{_{ij}}\rho _{_{ji}}\right) }.
\end{equation}
Although the concurrence and therefore the results we obtain are not
restricted to the standard one-mode three-level system, we will use that
language throughout most of the paper.

Having specified the various photonic crystal and field amplitude
parameters, we will present in the following subsection the results of our
numerical analysis of the concurrence.

\subsection{Numerical results}

For applications in real systems, we consider the dipole emitters with
frequencies in the reststrahl band of \ GaAs. In this subsection we will
discuss the time dependence of the concurrence, which considered as an
entanglement measure. We will consider the commonly used state as initial
condition for the cavity field: the coherent state, which may be applicable
in different situations. As might be expected, the behavior of the
three-level system changes dramatically depending on the initial field
state. Throughout this subsection the quantity to be examined is the
concurrence $C\left( \left\vert \psi \right\rangle \right) .$ 
\begin{figure}[tbph]
\begin{center}
\includegraphics[width=9cm]{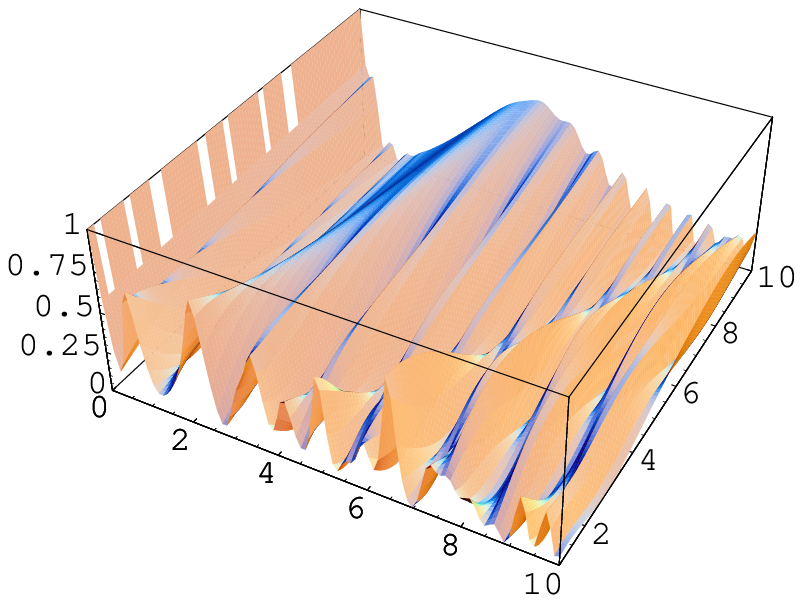} \includegraphics[width=9cm]{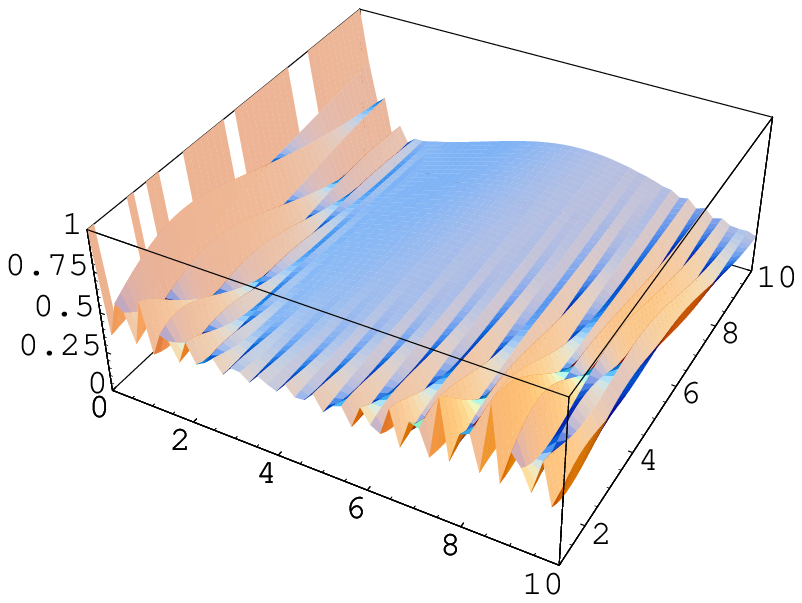}
\end{center}
\caption{The evolution of the concurrence $C\left( \left\vert \protect\psi %
\right\rangle \right) $ as a function of the scaled time $\protect\lambda %
_{1}t$ and the mean photon number $\bar{n}$. The parameters are $ \protect%
\epsilon _{0}=10.89,$ $\protect\eta =1.085,$ $\protect\omega /\protect\omega %
_{T}=2,$ $\protect\omega _{0}/\protect\omega _{T}=1,$ $\hslash \protect%
\omega _{L}=36.29$meV$,$ $\hslash \protect\omega _{T}=33.25$meV, $d_{1}=500$ 
\AA , $d_{2}=300$\AA , $\protect\epsilon _{1}=9,$ $\protect\epsilon %
_{2}=1.3, $ $d_{3}=500$\AA , $d_{4}=400$\AA , $\protect\epsilon _{3}=10,$ $ 
\protect\epsilon _{4}=1.5$ and $L=1.5d$, and different values of the
detuning parameter, where $\Delta =0$ for Fig. 3a and $\Delta =5\protect%
\lambda _{1}$ for Fig 3b. }
\end{figure}

In figure 3, we present the oscillatory behavior of the concurrence $
C\left( \left| \psi \right\rangle \right) $ against the scaled time $\lambda
_{1}t$ and the mean photon number $\overline{n}$ for different values of the
detuning parameter, where $\Delta =0$ for Fig. 3a and $\Delta =5\lambda _{1}$
for Fig 3b. We consider a specific system in which the cavity is taken as
GaAs with $\epsilon _{0}=10.89,$ $\eta =1.085,$ $\omega /\omega _{T}=2,$ $
\omega _{0}/\omega _{T}=1,$ $\hslash \omega _{L}=36.29$meV$,$ $\hslash
\omega _{T}=33.25$meV. The photonic crystals parameters are given by the
arbitrary set, $d_{1}=500$\AA , $d_{2}=300$\AA , $\epsilon _{1}=9,$ $
\epsilon _{2}=1.3,$ $d_{3}=500$\AA , $d_{4}=400$\AA , $\epsilon _{3}=10,$ $
\epsilon _{4}=1.5$ and $L=1.5d.$ The general behavior due to the coherent
state of the field does not contain any surprises it is quite broad,
corresponding to the standard quantum limit. The value of concurrence at the
first maximum is 1, which is quit remarkable, see figure 3a. After the time
goes on, we see that the maximum value of the concurrence decreases with
small amplitude of the oscillations. As the mean photon number increased,
the number of oscillations decreased.

The effect of the parameter which describes the mismatch between the atomic
frequency and the mean frequency of the cavity mode has been considered in
figure 3b. We set the other parameters as the same as in figure 3a, and $
\Delta =5\lambda _{1}$. As $\Delta $ is increased the behavior of the
three-level system becomes increasingly erratic. Shorter revival times cause
successive revivals to overlap and interfere so that the time evolution
appears irregular. The detuning parameter at which irregularity emerges is
closely tied to the mean-photon number: the higher the mean-photon number,
the smaller the detuning needed to produce irregular behavior. Larger
detuning also results in decreased revival amplitude due to the larger
number of frequencies in the sum, which causes the rephasing to be less
complete. However, a signature of the revivals persists as a return to the
bare Rabi frequency even at mean-photon number high enough that the behavior
looks random and the revival amplitude is essentially washed out. From our
further calculations (which are not presented here), we point out that as we
increase the value of the detuning one can see that the revival time is also
prolonged, however the period of fluctuations is decreasing. Detuning
affects the revival time by elongating it and the maximum value of the
entanglement degree becomes smaller and smaller. Similar to the case of a
two-level atom, detuning shifted the atomic occupation probability around
which it oscillates upward meaning that the energy is stored in the atomic
system.

\begin{figure}[tbph]
\begin{center}
\includegraphics[width=9cm]{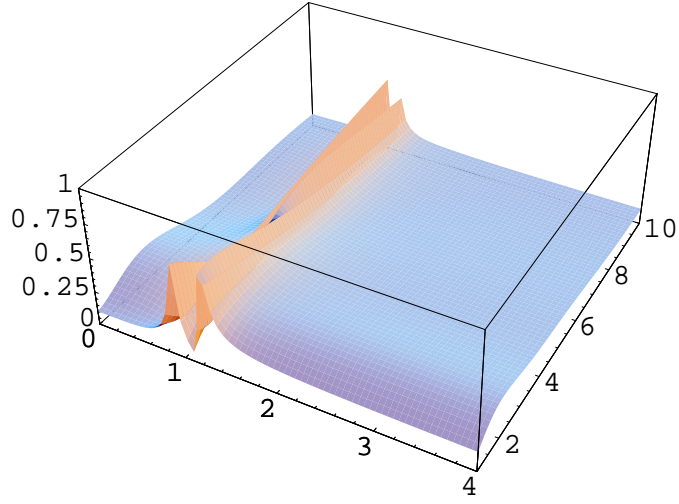} \includegraphics[width=9cm]{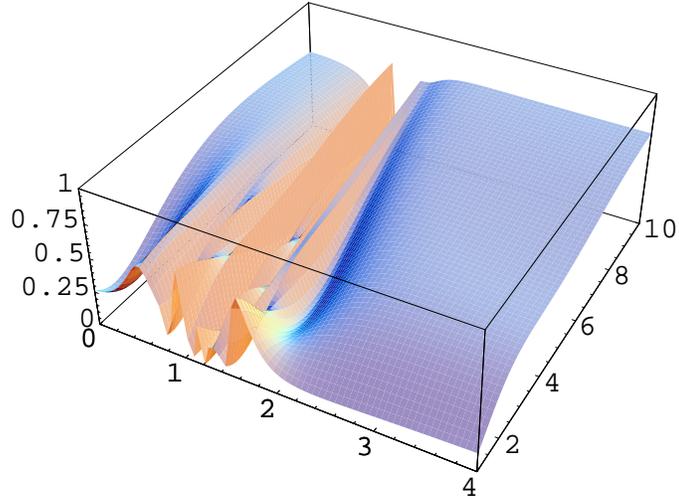}
\end{center}
\caption{The evolution of the concurrence $C\left(\left|\protect\psi %
\right\rangle\right)$ as a function of the mode-frequency $\protect\omega/%
\protect\omega_T$ and the mean photon number $\bar{n}$ for different values
of the scaled time, where $\protect\lambda_1t =\protect\pi/2$ for Fig. 4a
and $\protect\lambda_1t =3\protect\pi/2$ for Fig 4b. }
\end{figure}

Now we will turn our attention to the effect on the concurrence of the mode
frequency $\omega /\omega _{T}$. In particular, we consider $\epsilon
_{0}=10.89,$ $\eta =1.085,$ $\omega _{0}/\omega _{T}=1$ and for different
values of the scaled time, where $\lambda _{1}t=\pi /2$ for Fig. 4a and $
\lambda _{1}t=3\pi /2$ for Fig 4b. Our particular observation is the maximum
entanglement occurs near the band edges, which corresponds to $\omega
=1.085\omega _{T}$. Near the band edges the wave vector parallel to the
interface reaches its maximum value, and this corresponds to the first two
relatively small peaks around the point $1.085$. In the gap region or the
reststrahl region of the $GaAs$ system no electromagnetic fields can
propagate and coupling is therefore suppressed. The extra peaks around the
point $1.085$ are attributed to local field effects and can be understood
from looking at equation (7) where $\lambda _{ij}$ has a pole at $\eta
=\omega /\omega _{T}.$ One has to bear in mind that the above calculation
did not take into explicit account the spatial dependence of the coupling
parameters. Therefore, a more careful calculation would have to take into
account the nonstationary property of the present system. The above model
calculations suggest that physical parameters such as mode frequency,
mode-atom coupling and cavity dielectric have important effects on the
entanglement. One can see that the oscillations collapse after few Rabi
periods and after an interval of time in which the concurrence is constant,
the oscillations reappear again. This revival then collapses and a new
revival begins.

This behavior highlights once again the role of the functional form of the
modified Rabi frequencies in controlling the time evolution of the
concurrence. Rabi frequencies which obtained in the present model are
similar to that obtained from the standard three-level model but involving a
frequency-dependent dielectric function. An important point to keep in mind
when comparing the results presented here with results from the usual
three-level system in the absence of the photonic band gap is that: they are
give a different feature relative to the entanglement. This raises an
interesting question: can one use the present system in building quantum
logic gates? Calculations and detailed discussion of this issue will be
presented in a forthcoming paper.\bigskip \bigskip

\section{Phase entropy}

One of the most striking features of quantum mechanics is the property that
certain observable cannot simultaneously be assigned arbitrarily precise
values. This property does not compromise claims of completeness for the
theory, since it may consistently be asserted that such observable cannot
simultaneously be measured to an arbitrary accuracy \cite{bia75}. The
Shannon entropies associated with the photon number distribution $P_{m}$ and
phase probability distribution $P(\theta ,t),$ 
\begin{eqnarray}
P_{m} &=&\langle m|\rho (t)|m\rangle ,  \notag \\
P(\theta ,t) &=&\langle \theta |\rho (t)|\theta \rangle ,
\end{eqnarray}
where $|m\rangle $ is the Fock state and $|\theta \rangle $ is the phase
state, are given respectively by \cite{bia75}{\ } 
\begin{eqnarray}
R_{N} &=&-\sum\limits_{m=0}^{\infty }P_{m}\ln P_{m},  \notag \\
R_{\psi } &=&-\int\limits_{2\pi }\left( P(\theta ,t)\ln P(\theta ,t)\right)
d\theta .
\end{eqnarray}
The entropic uncertainty relations for the number and phase distribution
determine the lower bound on the sum of the Shannon entropies $R_{N}$ and $
R_{\psi }:$ 
\begin{equation}
R_{N}+R_{\psi }\geq \ln (2\pi ).
\end{equation}
This equality is satisfied by a Fock state for which $R_{N}=0$ and $R_{\psi
}=\ln (2\pi ).$ Other physical states give an entropic sum greater than $\ln
(2\pi ).$ Specifically, for a coherent state we find that the sum is $\ln
(e\pi )$ for the mean photon number greater than one, i.e. 
\begin{equation}
R_{N}+R_{\psi }\geq \ln (e\pi ).
\end{equation}
The lower bound for the position-momentum entropic uncertainty relation is
also given by right-hand side of this equation i.e $\ln (e\pi ).$

The single-mode of the Pegg-Barnett phase formalism which of interest in the
field of quantum optics can be constructed from the single-mode phases \cite%
{oba98} to take the form 
\begin{equation}
P(\theta ,t)=\lim_{s\rightarrow \infty }\biggl(\frac{s+1}{2\pi }\biggr) %
\langle \theta _{m}|\rho (t)|\theta _{m}\rangle ,  \label{31}
\end{equation}
$|\theta _{m}\rangle $ is a phase state of the mode, 
\begin{equation}
|\theta _{m}\rangle =\frac{1}{\sqrt{(s+1)}}\sum_{n=0}^{s}e^{in\theta
_{m}}|n\rangle ,  \label{32}
\end{equation}
where $\theta _{m}=\theta _{\circ }+\frac{2\pi m}{s+1},$ and $m=0,1,...s,$
and $\theta _{\circ }$ arbitrary. Equation (26) defines a particular basis
set of $(s+1)$ mutually orthogonal phase states.

\begin{figure}[tbph]
\begin{center}
\includegraphics[width=9cm]{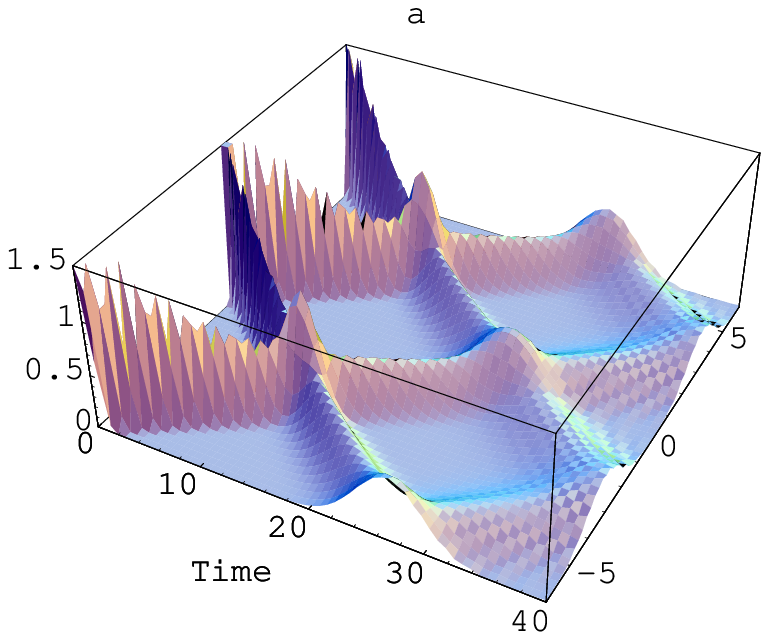} \includegraphics[width=9cm]{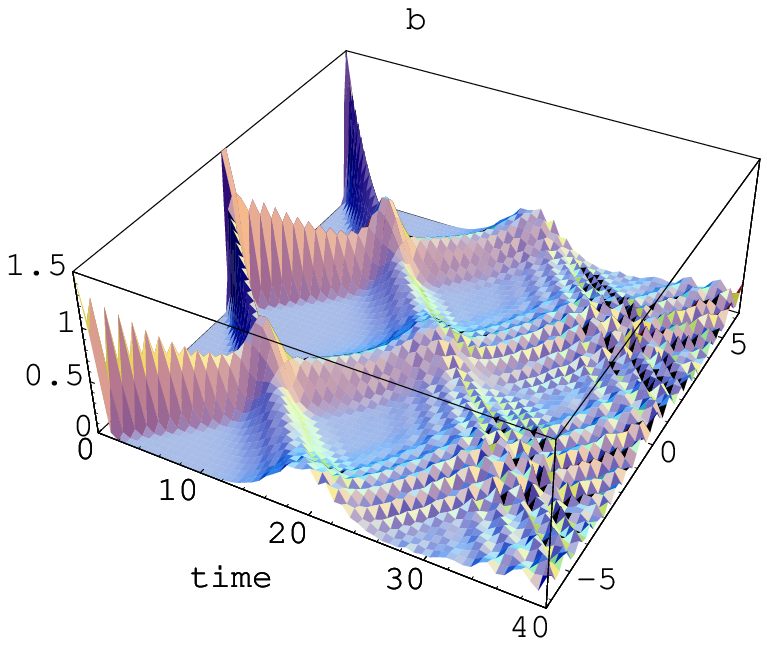}
\end{center}
\caption{$P(\protect\theta ,t)$ against $\protect\theta $ and the scaled
time $\protect\lambda _{1}t$. The parameters are $\protect\epsilon %
_{0}=10.89,$ $\protect\eta =1.085,$ $\protect\omega /\protect\omega _{T}=2,$ 
$\protect\omega _{0}/\protect\omega _{T}=1,$ $\hslash \protect\omega %
_{L}=36.29$meV$,$ $\hslash \protect\omega _{T}=33.25$meV, $d_{1}=500$\AA , $
d_{2}=300$\AA , $\protect\epsilon _{1}=9,$ $\protect\epsilon _{2}=1.3,$ $
d_{3}=500$\AA , $d_{4}=400$\AA , $\protect\epsilon _{3}=10,$ $ \protect%
\epsilon _{4}=1.5$ and $L=1.5d$, where (a) $\Delta =0$ and (b) $\Delta =5 
\protect\lambda _{1}.$ }
\end{figure}
Using the standard procedure \cite{oba98}, the phase probability
distribution, the expectation value and the variance of the Hermitian phase
operator may be obtained for the field. Since the coherent field at $t=0$
belongs to a class of partial phase states, we have chosen the reference
phase $\theta _{0}$ as $\theta _{0}=\beta -\frac{\pi s}{s+1},$ and
introduced the new phase labels $\zeta =m-\frac{1}{2}s$ where $
m=0,1,2,...,s. $ Then as $s$ tends to infinity the summation may be
transformed into an integral after replacing $\frac{2\pi \zeta }{s+1}$ by $
\theta ,$ and $\frac{2\pi }{s+1}$ by $d\theta .$ This leads to continuous
phase probability distribution, where 
\begin{equation}
P(\theta ,t)=\frac{1}{2\pi }\left( 1+2\sum\limits_{n>m}^{\infty }\left\{
A_{n,m}(t)\cos [\theta (n-m)]+B_{n,m}(t)\sin [\theta (n-m)]\right\} \right) ,
\end{equation}
where $A_{n,m}(t)$ and $B_{n,m}(t)$ are given by 
\begin{eqnarray}
A_{n,m}(t) &=&{Re}\left\{ A_{n}(t)A_{m}^{\ast }(t)+B_{n}(t)B_{m}^{\ast
}(t)+C_{n}(t)C_{m}^{\ast }(t)\right\} ,  \notag \\
B_{n,m}(t) &=&{Im}\left\{ A_{n}(t)A_{m}^{\ast }(t)+B_{n}(t)B_{m}^{\ast
}(t)+C_{n}(t)C_{m}^{\ast }(t)\right\} .
\end{eqnarray}
The phase probability distribution is normalized according to $\int_{-\pi
}^{\pi }P(\theta ,t)d\theta =1.$

\subsection{Numerical results}

In what follows we shall display some general arguments based on the
equality sign in the Heisenberg uncertainty relations that to demonstrate
the phase entropy of a general three-level system in the presence of
photonic band gab when the initial state of the field is assumed to be in a
coherent state.

In figure 5a, we have plotted the phase probability distribution $P(\theta
,t)$ as a function of the scaled time $\lambda _{1}t$ and $\theta $ taking
into consideration the presence of the photonic band gap. For example at
time $\lambda _{1}t=0$ we realize that the phase distribution $P(\theta ,t)$
starts with a single-peaked structure at $\theta =0$ corresponding to the
initial coherent state. Then as the time develops the peak splits into two
peaks moving into two opposite directions. 
\begin{figure}[tbph]
\begin{center}
\includegraphics[width=9cm]{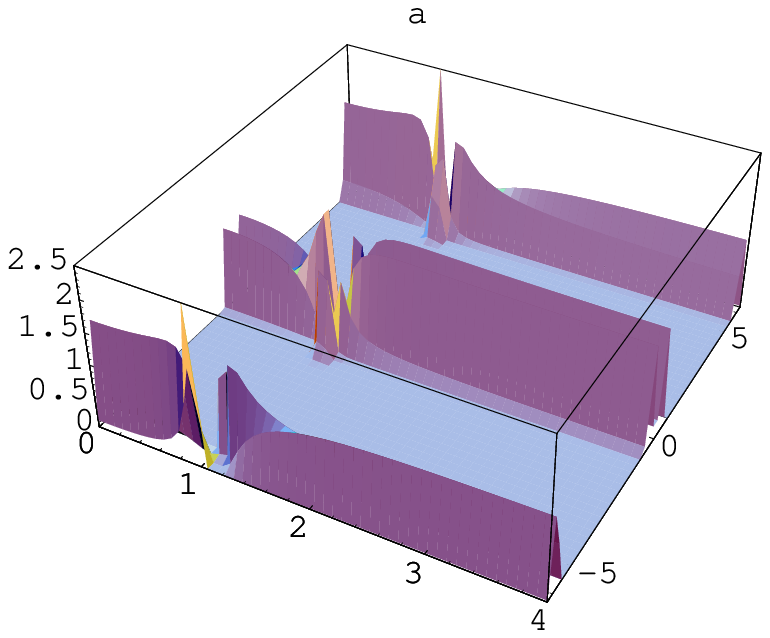} \includegraphics[width=9cm]{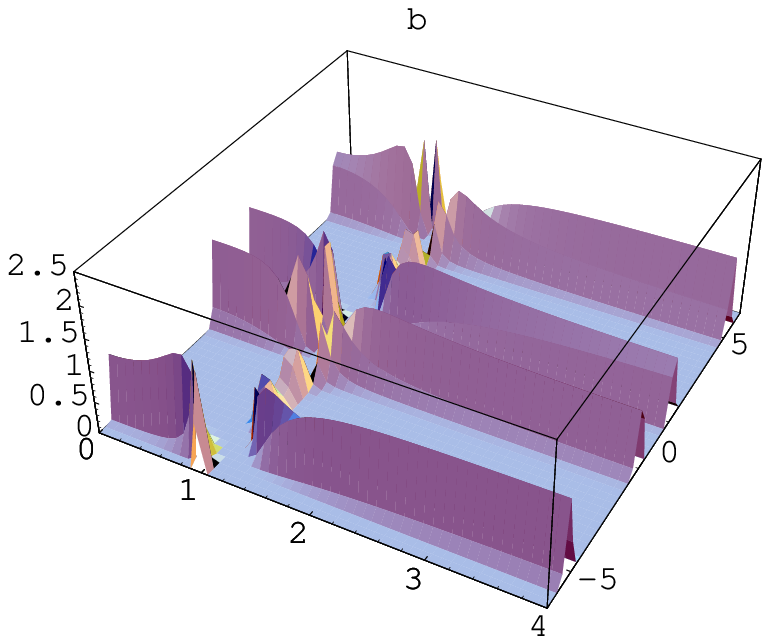}
\end{center}
\caption{$P(\protect\theta ,t)$ against the mode-frequency $\protect\omega /%
\protect\omega _{T}$ and $\protect\theta $. The parameters are the same as
in figure 5, but (a) $\protect\lambda _{1}t=\protect\pi /2$ and (b) $ 
\protect\lambda _{1}t=3\protect\pi /2.$ }
\end{figure}
However the amplitudes of the split peaks fluctuate in time giving a top
like shape until the two peaks reach the values $\theta =\pm \pi $ at middle
of the revival time but in this range the amplitudes of the peaks do not
show any fluctuations. The picture changes greatly as time develops further
(say $\lambda _{1}t>40$) where we find that the two-peak profile breaks up
into multi peak with reduction of the amplitudes of these peaks. Thus the
phase distribution shows diffusion as well as bifurcation. Different
features are visible when we consider the off-resonant case and the behavior
of the phase probability distribution is changed dramatically (see figure
5b). In this case we observe that there is a diffusion of the peaks at
earlier time.

In figure 6, we consider the behavior of the phase probability distribution
against the mode frequency $\omega /\omega _{T}$ and $\theta $ for the same
parameters as in figure 5, while in this figure, we keep the the scaled time 
$\lambda _{1}t$\ fixed, where, $\lambda _{1}t=\pi /2$ for figure 6a and $
\lambda _{1}t=3\pi /2$ for figure 6b. One may clearly see that the phase
probability distribution is discontinuous near the band edges. This
corresponds to the zero value at the point $1.085$. We can prove, in an
analogous manner to the equation (7), that the $\omega /\omega _{T}=1.085$
is a pole of the atom-field coupling\ can not avoided. It is interesting to
see that, the phase probability distribution does not depend on the mode
frequency for a fixed value of $\theta ,$ except at the point $1.085$. As
the time increased, the only difference is that, the phase probability
distribution peak splits into two peaks moving into two opposite directions,
keeping the symmetry around the point $\theta =0.$

\begin{figure}[tbph]
\begin{center}
\includegraphics[width=9cm]{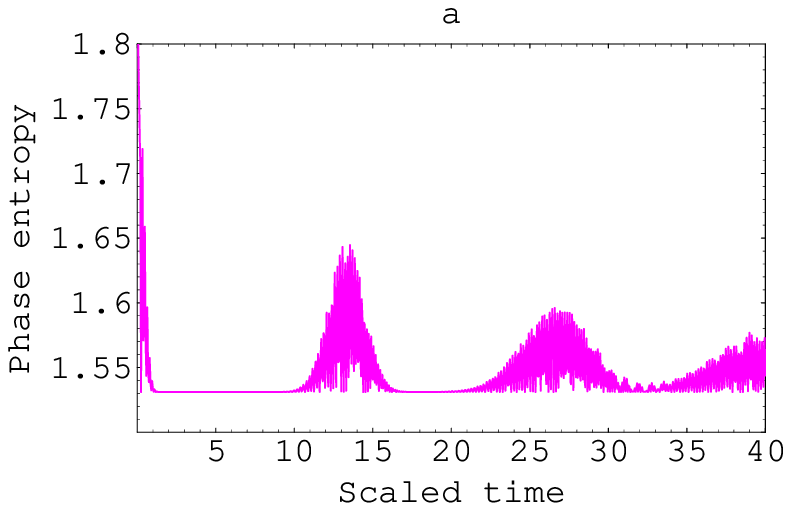} \includegraphics[width=9cm]{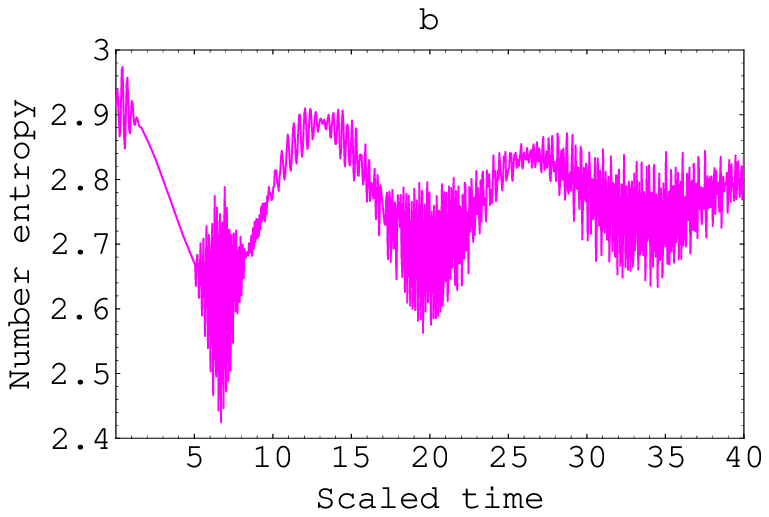}
\end{center}
\caption{The phase entropy $R_{\protect\psi }$ (a) and the number entropy $
R_{N}$ (b) as functions of the scaled time $\protect\lambda _{1}t$ with an
initial coherent state of the radiation field with $\bar{n}=20$ based on the
exact numerical results due to equation (23). We consider the same
parameters as in figure 1. }
\end{figure}
In figure 7, we plot the number entropy $R_{N}$ and the phase entropy $
R_{\psi }$ as functions of the scaled time $\lambda _{1}t$. The initial
state of the field is considered as a coherent state. We specifically
present the results for the same values of figure 5. It should be noted that
at a special choice of the mean-photon number parameter, the situation
becomes interesting, where the Rabi frequency has a minimum value at $\bar{n}
$. In this case we find that the general behavior of the entropies $R_{N}$
and $R_{\psi }$ and with an initially coherent field exhibit irregular
structures instead of the regular structure resembling those manifested by
the number or vacuum states cases. 
\begin{figure}[tbph]
\begin{center}
\includegraphics[width=9cm]{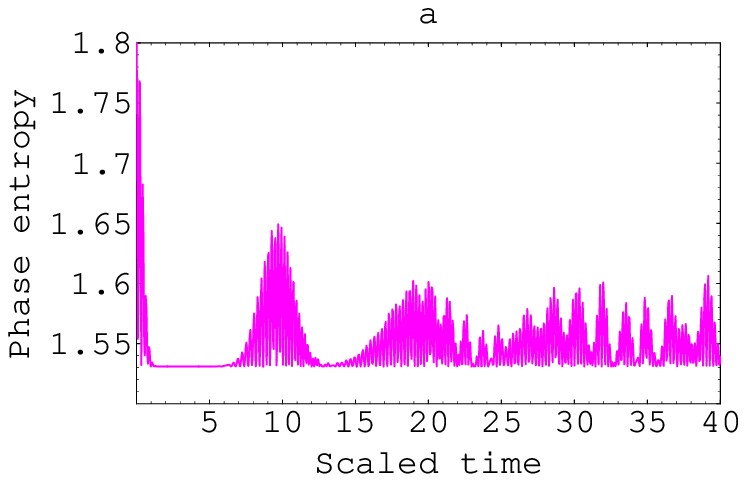} \includegraphics[width=9cm]{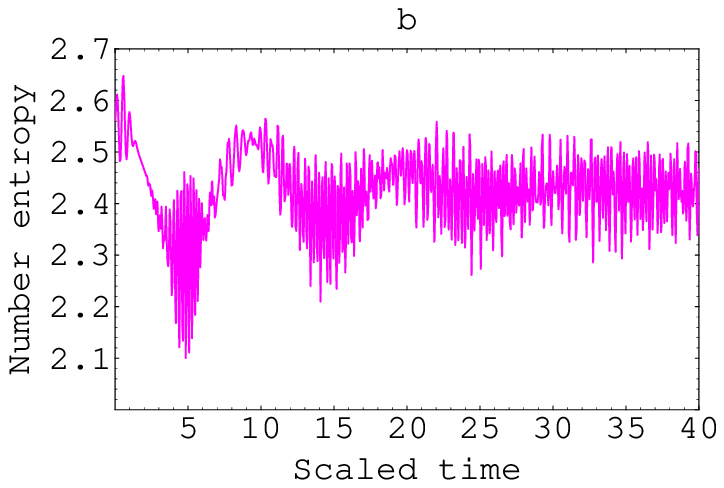}
\end{center}
\caption{The phase entropy $R_{\protect\psi }$ (a) and the number entropy $
R_{N}$ (b) as functions of the scaled time $\protect\lambda _{1}t$ where, $ 
\bar{n}=10$ and the same parameters as in figure 1. }
\end{figure}
Here it is interesting to note that the periodic oscillations are observed
for a short period of the interaction time only. When we consider smaller
mean-photon number, the regularity behavior of the oscillations in the
entropies $R_{N}$ and $R_{\psi }$ are still obvious (see figure 8) where we
have considered the initial mean photon number $\overline{n}=10$. However,
the number of oscillations is increased. Also it is interesting to point out
that at the revival time optimal phase entropy is attained in all the cases
which means that the atom has achieved an almost pure state, this has been
observed all through our figures. The number and phase entropic
uncertainties for a weak coherent state follow those of underlying states
superposition.

\begin{figure}[tbph]
\begin{center}
\includegraphics[width=9cm]{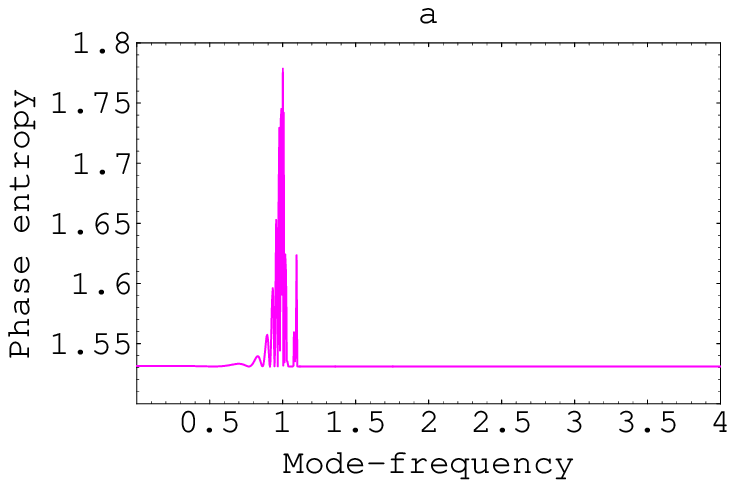} \includegraphics[width=9cm]{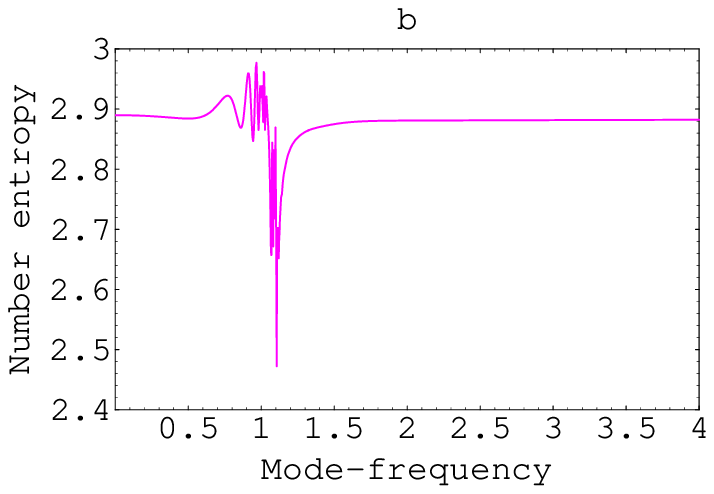}
\end{center}
\caption{The phase entropy $R_{\protect\psi }$ (a) and the number entropy $
R_{N}$ (b) as functions of the mode-frequency $\protect\omega / \protect%
\omega _{T}$, where, $\bar{n}=20, \protect\lambda_1t=\protect\pi /2$. }
\end{figure}

In figure 9 we plot the entropies $R_{N}$ and $R_{\psi }$ against the
mode-frequency $\omega $ in units of $\omega _{T}$ for different values of
the scaled time. Now, where the atom-field coupling is proportional to $
\lambda _{ij}$ this explains the origin of the second peak in this figure.
It is interesting to note the dependence of these entropies on the
mode-frequency, with different values of the scaled time. We wonder, as a
possible generalization of this concept, whether there exist another family
of similar oscillations if we consider two-qubit system. In such a case, the
properties of these systems would probably be of interest, in order to bring
further insight and knowledge about entanglement and quantum logic gates for
multi-partite systems. As the scaled time increased, a characteristic
feature of the entropies $R_{N}$ and $R_{\psi }$ is quit interesting, where
more oscillations exist, also, only around the resistable region for the
number entropy but less number of oscillations exist for the phase entropy
(see figure 10). 
\begin{figure}[tbph]
\begin{center}
\includegraphics[width=9cm]{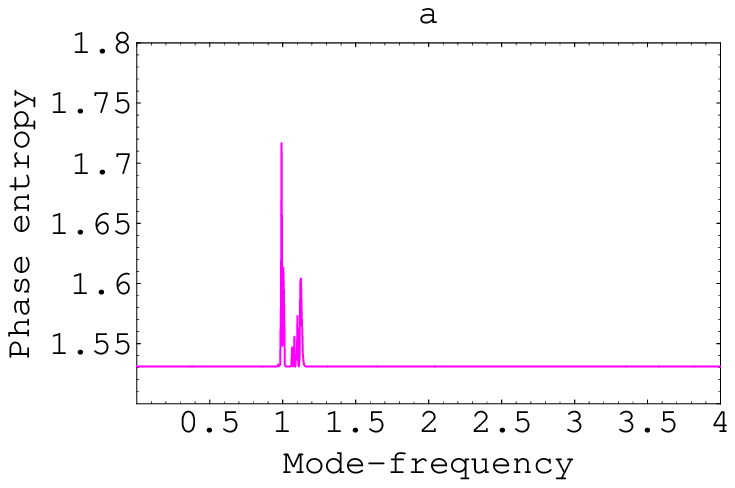} %
\includegraphics[width=9cm]{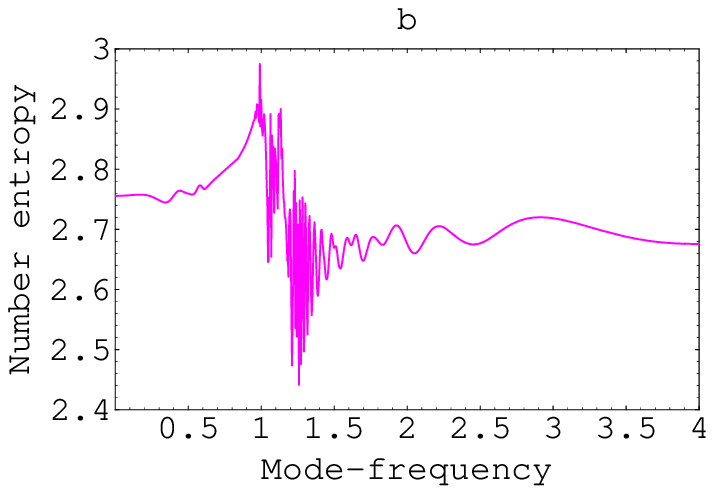}
\end{center}
\caption{The phase entropy $R_{\protect\psi }$ (a) and the number entropy $
R_{N}$ (b) as functions of the mode-frequency $\protect\omega / \protect%
\omega _{T}$, where, $\bar{n}=20,\protect\lambda _{1}t=3\protect\pi /2$. }
\end{figure}
Far from the resistable region the entropies behavior observed here does not
depend on the mode frequency and the intensity of the initial field mode.

Given a $3D$ photonic crystal with a complete gap, one has the possibility
of introducing a defect in the structure which will create a localized state
in the gap. If this is a point-like defect then the photon mode will be
completely localized about a point. In figure 10, we show the zero point
associated with the defect created by removing a small amount of dielectric
from one of the vertical dielectric columns of the crystal structure. The
resulting defect mode has a state near mid gap. One feature that should be
highlighted in this context is the appearance of a frequency gap between the
pair of interface dispersion. This gap is present only when the two photonic
crystal regions are different, and disappears when they are identical.

\section{Experimental prospects}

The perfect semiconductor crystal is quite elegant and beautiful, but it
becomes ever more useful when it is doped. Likewise, the perfect photonic
crystal can become of even greater value when a defect is introduced \cite%
{yab93}. The point to make about photonic crystals is that they are very
empty structures, consisting of about 78\ empty space. But in a sense they
are much emptier than that. They are emptier and quieter than even the
vacuum, since they contain not even zero-point fluctuations within the
forbidden frequency band. Our model system consist of a three-level atom
located inside a photonic band gap material. There are several ways of
placing such an atom inside a photonic crystal. From a material standpoint,
it is possible to dope an existing photonic band gap material using ion beam
implantation methods. For instance, it has recently been shown that $Er^{3+}$
ions implanted into bulk silicon exhibit sharp free-atom-like spectra \cite%
{yab93,lan98}. Intense temperature-dependent photoluminescent (PL) at $1.54{%
\ \mu }m$ is observed in the system at low temperatures (when the host
material is crystalline, Er-related PL is quenched at temperatures above $
80K $ so that it cannot be detected at room temperatures). This wavelength
is particularly significant because it corresponds to the minimum absorption
of silica fibre-based optical communication system. Because the PL at $
1.54\mu m $ is due to the spin-orbit split $^{4}I_{13/2}\rightarrow
^{4}I_{15/2}$ of $4f$ electrons in the $Er^{3+}$ ions which are shielded by
outer $5s^{2}5p^{6} $ shells, the influence of the host lattice on the
luminescence wavelength is weak. (The key to the success of erbium is that
the upper level of the amplifying transition $^{4}I_{13/2}$ is separated by
a large energy gap from the next-lowest level $^{4}I_{15/2}$ so that its
lifetime is very long and mostly radiative. In spite of the screening of the
atomic transition by the outer shells, it is likely that thermal phonons in
the silicon host would cause significant dephasing of the quantum degrees of
freedom within the erbium 4f shell. Consequently, such a system must be
cooled to liquid helium temperatures. Such experiments appear to be nearly
within the reach of current technology. Although it has not yet been
demonstrated, the system consisting of a multi-level system coupled to a
multi-mode appears to be another potential candidate for achieving new
features. {Such systems are potentially interesting for their ability to
process information in a novel way and might find application in models of
quantum logic gates. Therefore, atoms or trapped ions + cavities in a
presence of photonic band gap represent, in our opinion, a very promising
system for quantum information processing.}

\section{Conclusion}

In this communication the quantum electrodynamic properties of a three-level
atom embedded in a photonic band gap material were investigated. We have
focused on the application of the effective-medium theory to the present
problem in a nanoscale dielectric cavity QED situation. The effective-medium
approach can in fact be applied to situations in which all three regions of
the structure possess frequency-dependent dielectric functions.
Specifically, the combined effects of coherent control by an external
driving field and photon localization facilitated by a photonic band gap on
entanglement from a three-level atom embedded in a photonic band gap
material were examined. Exact solutions of the wave function in the Schr\"{o}%
dinger picture have been obtained within rotating wave approximation. In
particular, we have chosen to focus on three-level system coupled to a
single mode. Observation of the three-level system may offer some insight
into the quantum nature of the resonator, just as atoms provide a sensitive
probe for the nonclassical nature of electromagnetic fields. The observation
of revivals, which are a strictly nonclassical phenomenon, would give
evidence for the quantum nature of the quantum system.

The results point to a number of interesting features, which arise from the
variation of the adjustable parameters of the system, namely, the
mode-frequency, dipole vector orientation, dipole position within the slab,
the slab width, and the photonic crystal parameters: layer widths and
dielectric functions. Our investigations for the entanglement,
collapse-revival phenomena, and phase and number entropic uncertainty
relations in the presence of the photonic band gap as compared with the
usual three-level model are summarized as follows:-

i) The concurrence behavior is reflect the pattern of collapse and revival
which is qualitatively similar to that of the usual three-level model but
with reduced amplitude. In case of a smaller mean photon number and for
initially excited atom the usual pattern in the three-level model of
collapse and revival changes to rapid fluctuations of interference patterns
for all time considered. In this way, our concurrence function contains all
the information necessary to identify the entanglement of a given state.
Nevertheless, it depends on the particular choice of the mode-frequency.

ii) The phase entropy can be used to measure entanglement of the system
presented here with explicitly atom-field coupling in the presence of
photonic band gap. We would like to point out that the phase Shannon
entropic considered for the presented model has not been treated in this
manner before.

iii) The photonic band gap introduces sudden changes in the concurrence and
phase entropy due to the variation of these quantities with mode frequency.
This feature attributed to the fact that in the photonic band gap region
electromagnetic modes are not allowed to propagate into the dielectric slab
and hence no interaction can take place in this region. Theory predicts
analytically this behavior for a GaAs system at $\omega =\eta \omega _{T}$.

Finally, we emphasize the fact that without any conditions it was possible
to obtain exact analytic solution which reproduce the most important
features of the three-level atom interacting with a cavity one- or two-mode
in the presence of photonic band gap. A similar set of equations have been
derived in \cite{bou04} for a three-level system using some approximations,
based on the Riccati nonlinear differential equation. In contrast, the
method used here gives exact analytic solutions without any conditions.

\newpage \textbf{Acknowledgment}

I acknowledge the hospitality and financial support from the Center for
Computational and Theoretical Sciences, Kulliyyah of Science, IIUM, Malaysia
where the final version of the paper was prepared. Also, helpful discussions
with Prof. A.-S. F. Obada and Prof. M. R. B. Wahiddin are gratefully
acknowledged.


\begin{thebibliography}{99}
\bibitem{lee04} R.-K. Lee, Y. Lai: J. Opt. B: Quantum Semiclass. Opt. 
\textbf{6}, S715 (2004); Jan Perina Jr., C. Sibilia, D. Tricca, M.
Bertolotti: Phys. Rev. A \textbf{70}, 043816 (2004)

\bibitem{yam03} S. Yamada, Y. Watanabe, Y. Katayama, J. B. Cole: J. Appl.
Phys. \textbf{93}, 1859 (2003); 5. S. Yamada, Y. Watanabe, Y. Katayama, X.
Y. Yan, J. B. Cole: J. Appl. Phys. \textbf{92}, 1181 (2002).

\bibitem{kam00} A. Kamli, M. Babiker: Phys. Rev A \textbf{62,} 043804 (2000).

\bibitem{joa95} J. D. Joannoupoulos, R. B. Meade, J. N. Winn: Photonic
Crystals: Molding the Flow of Light (Princeton University Press, Princeton
N.J., 1995); S. John: Phys. Rev. Lett. \textbf{58} 2486 (1987); E.
Yablonovitch: Phys. Rev. Lett. \textbf{58,} 2059 (1987).

\bibitem{joh90} S. John, J. Wang: Phys. Rev. Lett. \textbf{64}, 2418 (1990).

\bibitem{joh94} S. John, T. Quang: Phys. Rev. A \textbf{50}, 1764 (1994).

\bibitem{zhu97} S.-Y. Zhu, H. Chen, H. Huang: Phys. Rev. Lett. \textbf{79},
205 (1997).

\bibitem{pas98} E. Paspalakis, P.L. Knight: Phys. Rev. Lett. \textbf{81},
293 (1998)

\bibitem{pas99} E. Paspalakis, N. J. Kylstra, P. L. Knight: Phys. Rev. A 
\textbf{60}, R33 (1999).

\bibitem{flo01} M. Florescu, S. John: Phys. Rev. A \textbf{64}, 033801
(2001).

\bibitem{vid03} G. Vidal, J. I. Latorre, E. Rico, A. Kitaev: Phys. Rev.
Lett. \textbf{90}, 227902 (2003); T. J. Osborne, M. A. Nielsen: Phys. Rev. A 
\textbf{66}, 032110 (2002); A. Osterloh, L. Amico, G. Falci, R. Fazio:
Nature \textbf{416}, 608 (2002); I. Bose, E. Chattopadhyay: Phys. Rev. A 
\textbf{66}, 062320 (2002).

\bibitem{buz98} V. Buzek, G. Drobny, M. S. Kim, G. Adam, P. L. Knight: Phys.
Rev. A \textbf{56}, 2352 (1997); X. Luo, X. Zhu, Y. Wu, M. Feng, K. Gao,
Phys. Lett. A \textbf{237}, 354 (1998); A. S. Parkins, H. J. Kimble: J.Opt.
B: Quantum Semiclass. Opt. \textbf{1} , 496 (1999); A. S. Parkins, E.
Larsabal: Phys. Rev. A \textbf{63}, 012304 (2000).

\bibitem{doh02} A. C. Doherty, P. A. Parrilo, F. M. Spedalieri: Phys. Rev.
Lett. \textbf{88}, 187904 (2002); O. Rudolph, quant-ph/0202121; M.
Horodecki, P. Horodecki, Horodecki: quant-ph/0206008; for a review see D.
Bru \ss\ et. al.: J. Mod. Opt. \textbf{49}, 1399 (2002).

\bibitem{per99} A. Peres: Found. Phys. \textbf{29}, 589 (1999).

\bibitem{hor96} M. Horodecki, P. Horodecki, R. Horodecki: Phys. Lett. A 
\textbf{223}, 1 (1996); B.M. Terhal: ibid. \textbf{271}, 319 (2000); M.
Lewenstein et al.: Phys. Rev. A \textbf{62}, 052310 (2000).

\bibitem{woo98} W. K. Wootters: Phys. Rev. Lett. \textbf{80}, 2245 (1998);
F. Verstraete, K. Audenaert, J.Dehaene, B. De Moor: J. Phys. A: Math. Gen. 
\textbf{34, }10327 (2001); S. Hill, W. Wootters: Phys. Rev. Lett. \textbf{80}
, 2245 (1998); Opt. Comm.\textbf{152}, 119 (1998); J. Math. Phys. \textbf{39}
, 4604 (1998).

\bibitem{bia75} I. Bialynicki-Birula, J. Mycielski: Commun. Math. Phys. 
\textbf{44}, 129 (1975); D. L. Deutsch, Phys. Rev. Lett. \textbf{50,} 631
(1983); H. Maassen, J. B. M. Uffink: Phys. Rev. Lett. \textbf{60,} 1103
(1988).

\bibitem{ben96} J. M. Bendickson, J. P. Dowling: Phys. Rev. E \textbf{53},
4107 (1996).

\bibitem{cor99} C. M. Cornelius, J. P. Dowling: Phys. Rev. A \textbf{59},
4736 (1999); A. J. Stimpson, J. P. Dowling, Thermal emissivity of 3d
photonic band-gap structures, presented at the Optical Society of American
Annual Meeting, Long Beach, CA (2001).

\bibitem{lin00} S.-Y. Lin, J. G. Fleming, E. Chow, J. Bur: Phys. Rev. B 
\textbf{62}, R2243 (2000).

\bibitem{joh93} S. John, N. Akozbek: Phys. Rev. Lett. \textbf{71}, 1168
(1993); Phys. Rev. E 57, 2287 (1998).

\bibitem{suk99} A.A. Sukhorukov, Yu.S. Kivshar, O. Bang: Phys. Rev. E 
\textbf{60}, R41 (1999).

\bibitem{mcg99} A. R. McGurn: Phys. Lett. A \textbf{251}, 322 (1999); Phys.
Lett. A \textbf{260}, 314 (1999).

\bibitem{cot89} M. Cottam, D. R. Tilley, Introduction to Surface and
Superlattice Excitations (Cambridge University Press, Cambridge, England,
1989); V. M. Agranovich, D. L. Mills, Surface Polaritons (North-Holland,
Amsterdam, 1982).

\bibitem{kon75} J. A. Kong: Theory of Electromagnetic Waves (Wiley, New
York, 1975).

\bibitem{zol98} F. Zolla, D. Felbacq, B. Guizal: Opt. Commum. \textbf{148},
6 (1998)

\bibitem{abd87} A. M. Abdel-Hafez, A.-S. F. Obada, M. M. A. Ahmed: Physica
A, \textbf{144}, 530 (1987); \ A. M. Abdel-Hafez, A. M. M. Abu-Sitta, A.-S.
F. Obada: Physica A, \textbf{156}, 689 (1989); J. H. McGuire, K. K. Shakov,
K. Y. Rakhimov: J. Phys. B \textbf{36}, 3145 (2003)

\bibitem{bou04} S. Bougouffa, A. Kamli: J. Opt. B: Quantum Semiclass. Opt. 
\textbf{6}, S60 (2004).

\bibitem{ben96} C. H. Bennett, D. P. DiVincenzo, J. A. Smolin, W.K.
Wootters: Phys. Rev. A \textbf{54}, 3824 (1996); C. H. Bennett, S. J.
Wiesner: Phys. Rev. Lett. \textbf{69}, 2881 (1992); C. H. Bennett, G.
Brassard, C. Crepeau, Rjozsa, A. Peres, W. K. Wootters: Phys. Rev. Lett. 
\textbf{70}, 1895 (1993).

\bibitem{run01} P. Rungta, V. Buzek, C. M. Caves, M. Hillery, G. J. Milburn:
Phys. Rev. A \textbf{64}, 042315 (2001); A. Lozinski, A. Buchleitner, K.
Zyczkowski, T. Wellens: Europhys. Lett. 62, 168 (2003).

\bibitem{oba98} A.-S. F. Obada, A. M. Abdel-hafez, M. Abdel-Aty: Eur. Phys.
J. D \textbf{3}, 289 (1998); D. T. Pegg, S. M. Barnett: Phys. Rev. A \textbf{%
\ 39}, 1065 (1989).

\bibitem{yab93} E Yablonovitch: J. Phys.: Condens. Matter \textbf{5}, 2443
(1993); M. Woldeyohannes, S. John: J. Opt. B: Quantum Semiclass. Opt. 
\textbf{5, }R43 (2003); C. M. Bowden, J. P. Dowling, H. O. Everitt: J. Opt.
Soc. Am. B \textbf{10}, 280 (1993); G. Kurizki, J. W. Haus: Special issue on
photonic band structures, J. Mod. Opt. \textbf{41}, 171 (1994).

\bibitem{lan98} S. Lanzerstorfer, L. Palmetshofer, W. Jantsch, J. Stimmer:
Appl. Phys. Lett. \textbf{72}, 809 (1998); X. Zhao, S. Komuro, H. Isshiki,
Y. Aoyagi, T. Sugano: Appl. Phys. Lett. \textbf{74} 120 (1999).
\end{thebibliography}
\end{document}